\documentclass[twocolumn,aps,floatfix,showpacs,tightenlines,superscriptaddress,amsmath,amssymb]{revtex4}
\pdfoutput=1
\usepackage{amsthm,amsbsy,epsfig,color,graphicx,times,bbm,enumitem}
\usepackage[ansinew]{inputenc}
\usepackage[english]{babel}
\usepackage{accents}
\usepackage{braket}
\usepackage{amsfonts}
\usepackage{booktabs}
\usepackage{tabularx}
\usepackage{array}
\usepackage{hyperref}
\hypersetup{pdfstartview=FitH}
\usepackage{latexsym}
\usepackage{multirow}
\usepackage{longtable}

\begin{document}

\title{Strong deformations of DNA: Effect on the persistence length}

\author{Kyry\l{}o A. Simonov}
\affiliation{Fakult\"{a}t f\"{u}r Mathematik, Universit\"{a}t Wien, Oskar-Morgenstern-Platz 1, 1090 Vienna, Austria}

%\date{\today}
\pacs{36.20.Hb, 87.14.Gk, 87.15.La}

\begin{abstract}
Extreme deformations of the DNA double helix attracted a lot of attention during the past decades. Particularly, the determination of the persistence length of DNA with extreme local disruptions, or kinks, has become a crucial problem in the studies of many important biological processes. In this paper we review an approach to calculate the persistence length of the double helix by taking into account the formation of kinks of arbitrary configuration. The reviewed approach improves the Kratky--Porod model to determine the type and nature of kinks that occur in the double helix, by measuring a reduction of the persistence length of the kinkable DNA.  %
\end{abstract}

\maketitle

\section{Introduction}

Many biological functions are intimately connected to the conformational deformability of DNA which can essentially influence its genetic activity. Numerous physical experiments and computer simulations of DNA have demonstrated its noticeable flexibility, particularly, certain proteins can cause a formation of the localised extreme bends (kinks) in the DNA structure~\cite{prevost}. Furthermore, the formation of extreme bends in the DNA structure could be caused by a wide variety of biological processes, such as intercalation of small molecules~\cite{saenger}, DNA conformational changes~\cite{chen}, packaging~\cite{richmond} and others.

The interest in the extreme bending of the double helix was initiated by Crick and Klug who investigated the mechanism of folding of DNA in chromatin, the chromosomal material of the cell nucleus. They first suggested that DNA is folded due to the formation of kinks in the double helix~\cite{crick}, and later such defects were found within experiments with a packaged DNA and DNA-protein complexes~\cite{richmond,dickerson,werner,suzuki}. The recent view on the problem of strong bending and kink formation in the double helix was refreshed by Cloutier and Widom who found that the short DNA fragments of 94 base pairs cyclise much more easily than one would expect from the theory~\cite{cloutier}. This crucial result caused intense discussion on the cyclisation of DNA. A possible explanation for the observed phenomenon was that localised distortions (kinks) in the DNA fragment lead to the formation of a sharp bend and an increase the probability of looping~\cite{yan,du}. Later, numerous computational experiments with DNA minicircles revealed the presence of various types of kinks~\cite{prevost,lankas,kotlyar,mitchell}.

Idealised models of chain molecules are suitable tools commonly used to describe configurations of the DNA chain~\cite{grosberg} and accordingly its deformations. Particularly, the Kratky--Porod model (and its continuous version --- the worm-like chain, WLC)~\cite{kratky} is considered as a basic mechanical model of DNA which can well describe a wide range of its mechanical properties~\cite{vologodskii}. This model considers a coil of a smoothly curved strand, the direction of curvature at any point of the strand being random~\cite{cantor,flory}. The Kratky--Porod chain carries several configurational parameters, which characterise the flexibility of the DNA coil in solution and can be measured in a hydrodynamic experiment. Particularly, the stiffness of DNA is determined by the persistence length $A$, which is a measure of distance over which the DNA chain 'remembers' the direction of the first segment. Hence, the directional correlation of two segments decreases exponentially with a typical length $A$ while increasing the contour length separating them~\cite{grosberg,cantor,bustamante}. However, it should be noted that the chains of the contour length equal to $A$ do not necessarily have a rigid-rod-like behaviour~\cite{hagerman}.

The Kratky--Porod model assumes that the energy cost of bending is a quadratic function of the bending angle~\cite{vologodskii}. Therefore, the Kratky--Porod model is expected to describe only the smooth deformations of the double helix with relatively small changes between the bonds. In this way, this model should be improved to include the configurations of the double helix with kinks of various nature, since corresponding approaches are still not completely developed. Over the past decades several ways for a renormalisation of the persistence length were proposed~\cite{popov,kulic}. Furthermore, Wiggins and colleagues proposed an extension of the WLC model --- the kinkable WLC model (KWLC), which includes sharp kinks characterised by a probability of such a kink occurring per unit length~\cite{wiggins}. However, these approaches consider the kinks to be of one type only, which is rather idealised. In~\cite{simonov} the Kratky--Porod chain was extended to a chain which can undergo the formation of kinks of different length and configuration.

The paper is organised as follows. In Section~\ref{sec:KinksWLC} we review the classification of kinks used in literature and proposed by Lanka\v{s} \textit{et al}~\cite{lankas} and generalise it to the case of kinks with more complicated structure. Then we review the approach presented in~\cite{simonov}, particularly discussing the way to distinguish normal and kinked states of a helix step and considering several configurational parameters that describe the flexibility of DNA. In Section~\ref{sec:Cations} we analyse the obtained approach and compare it with the approach for the bending of DNA by multivalent cations proposed by Rouzina and Bloomfield~\cite{rouzina}. Last but not least we present the summary and outlook in Section~\ref{sec:Summary}.

\section{Kinks in the Kratky--Porod model}
\label{sec:KinksWLC}

Sharp structural changes of the double helix could cause the breaking of base-stacking interactions and hydrogen bonds. Particularly, the main property of kinks is that their formation is specified by the break of the stacking between consecutive base pairs. Moreover, different biological processes cause kinks of different structure and hence different chan\-ges of stacking. In this way, kinks in the double helix can be classified by their influence on base-stacking interactions.

\begin{figure}
\centering\resizebox{0.3\textwidth}{!}{%
  \includegraphics{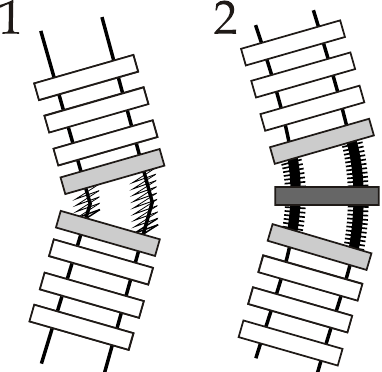}
}
\caption{Diagrammatic representation of the two types of kinks in the double helix: 1) type 1 --- loss of stacking in one helix step (involving two intact base pairs), 2) type 2 --- changing of stacking in two helix steps (involving a modified central base pair and two intact outer base pairs).}
\label{KinkTypes}
\end{figure}

Computer simulations of the minicircles of 94 base pairs detected two basic groups of kinks with different structures leading to the strong bends towards the major groove~\cite{lankas,maher,harris}. As shown schematically in Fig.~\ref{KinkTypes}, a \textit{type 1} kink is caused by unstacking a single helix step, whereas a \textit{type 2} kink involves two helix steps where the central pair is claimed to be broken. However, we extend the definition of a type 2 kink by using the more general case of a \textit{modified} central base pair to include more factors which could cause the formation of such kinks.

A \textit{type 1} kink is an extreme local deformation in one helix step which looks like a kink proposed by Crick and Klug~\cite{crick,curuksu}. Originally Crick and Klug described a sharp kink of $98^{\circ}$ involving two base pairs with disrupted stacking that causes a strong bend of the double helix~\cite{crick}. Kinks as extreme as Crick--Klug kinks are rather idealised, however the studies of the nicked double helix showed that the formation of similar kinks in the sites of single-stranded breaks is energetically favorable~\cite{protozanova,yakovchuk}. The molecular dynamics simulations of the series of DNA minicircles also showed the formation of type 1 kinks in some cases, particularly for the minicircles containing 64--66 base pairs~\cite{kotlyar,mitchell}.

Furthermore, similar damages of the DNA structure can be caused by intercalation of small molecules into the double helix, for example due to the sequence-dependent binding of a protein to DNA~\cite{chen,saenger,dickerson,werner,suzuki}. In particular, the studies of the binding of $\Delta$-[Ru$($phen)$_3$]$^{2+}$~\cite{reymer} and \newline [Ru(TAP)$_2$(dppz)]$^{2+}$~\cite{hall} complexes to DNA showed that the intercalating ligands act as wedges in the minor groove, thereby inducing type 1 kinks in the double helix.

\textit{Type 2} represents a kink with the changed stacking induced by a modification of a base pair. Such kinks are distributed over two base-pair steps in this case, with a modified central base pair and maintained outer base pairs~\cite{prevost,lankas,curuksu}, as shown in Fig.~\ref{KinkTypes}. Such a structure is treated as more probable to appear in the double helix than kinks of type 1~\cite{prevost}. For example, computer simulations of DNA oligomers of 15 base pairs detected the presence of type 2 kinks only~\cite{curuksu}. In the simulations of the DNA minicircles containing 65 and 110 base pairs almost all detected kinks had a type 2 structure except for the case of relaxed 65-base pair minicircle which experienced also a type 1 kink between d(GC) base pairs~\cite{mitchell}.

An opened base pair should provide high local flexibility~\cite{vologodskii}, therefore it is a natural candidate to cause the formation of a kink holding the structure of a type 2 kink. The occurrence of the partially opened (preopened) configuratios of a base pair with possible binding of a water molecule to it~\cite{volkov1995,kryachko} could be a reason to induce a type 2 kink in the double helix as well. Besides, the discovery of the formation of Hoogsteen base pairs in a linear DNA with a finite probability, which could provide high local flexibility as well~\cite{vologodskii,nikolova}, reveals another candidate to cause a type 2 kink.

\begin{figure}\centering\resizebox{0.35\textwidth}{!}{%
  \includegraphics{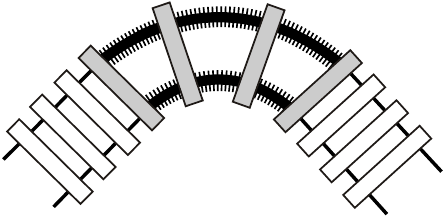}
}\caption{Diagrammatic representation of the particular conformational kink distributed over 4 intact base pairs in the double helix.}\label{ConfKink}\end{figure}

Consequently, in the literature one distinguishes two types of kinks with specified configurations. However, the geometry of kinks is not known for all the structural changes in the double helix. In particular, B--A transformations~\cite{volkov,lu,olson}), binding of the TBP to TATA-box~\cite{klug,kanevska} and the presence of A-tracts and GGCC-tracts~\cite{murphy,hanlindsay,handlakic} could cause the formation of kinks holding a more complicated structure than type 1 and type 2 kinks have. For example, TBP induces a strong composite bend over 8 base pairs of TATA-box which includes sharp bends (kinks) by $52^{\circ}$ and $39^{\circ}$ at the first and last base pairs, and smoother bend at $90^{\circ}$ within 6 central base pairs~\cite{kim,nikolov}. Therefore, we can extend the notion of a kink and take into account the formation of smoother kinks which involve four or more maintained base pairs with changed stacking~\cite{volkov_private}. The distribution of the bending angles inside such kinks is arbitrary. In that way, we will further consider kinks distributed over $n\geq 3$ helix steps with conformational changes and refer to such kinks as \textit{conformational}. Fig.~\ref{ConfKink} shows a particular smooth conformational kink which involves four base pairs (so 3 helix steps).

Summing up, we will further use the following classification of kinks in the double helix,
\begin{itemize}
 \item \textit{type 1 kink}, involving $n=1$ helix step,
 \item \textit{type 2 kink}, involving $n=2$ helix steps,
 \item \textit{conformational kink}, involving $n\geq 3$ helix steps.
\end{itemize}

\begin{figure}[h!]\centering\resizebox{0.3\textwidth}{!}{%
  \includegraphics{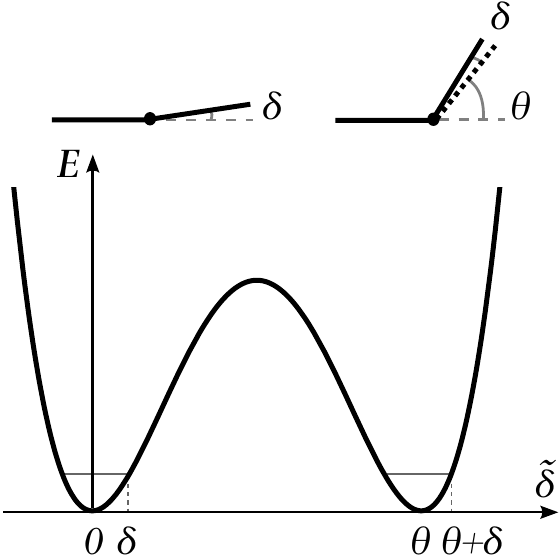}
}
\caption{Top: two possible states of a monomeric element, a normal one (with the bending angle $\delta$) and a kinked one (with the bending angle $\theta + \delta$). Bottom: the minima of the double well represent two states of a monomeric element.}\label{DoublePotential}\end{figure}
Each kink in the double helix is determined by its total angle $\theta$ and the phenomenological probability $W$ of this kink occurring. Indeed, this probability corresponds to the concentration of kinks in the double helix and can depend on several parameters, e.g. bending energy $E$, spring constant $k$, etc. Particularly, in the KWLC model $W = 2ke^{-E}$ for kinks of type 1~\cite{wiggins}.

As we mentioned above, the bending energy $E$ in the Kratky--Porod model depends quadratically on the small bending angle $\delta$. Since we aim to take into account the formation of kinks involving extreme angles, we turn to a double well potential~\cite{simonov,krumhansl}, which represents two possible states of a monomeric element, a normal and a kinked one, as shown in Fig.~\ref{DoublePotential}. The monomeric element in the normal state is bent at the small angle $\delta$ due to the thermal fluctuations just as in the Kratky--Porod model. The kinked state of a monomeric element describes the presence of a kink with a certain angle $\theta$. Therefore, we can represent the total bending angle $\tilde{\delta}$ of a monomeric element in a kinked state as the sum of a kink angle ($\theta$) and a small deviation ($\delta$)~\cite{simonov},
\begin{eqnarray}
\nonumber
& \tilde{\delta} = \theta \pm \delta,\\
& \cos \tilde{\delta} = \cos \theta \cos \delta \mp \sin \theta \sin \delta \approx \cos \theta \cos \delta.
\end{eqnarray}

\begin{figure}[h!]
\centering\resizebox{0.45\textwidth}{!}{%
  \includegraphics{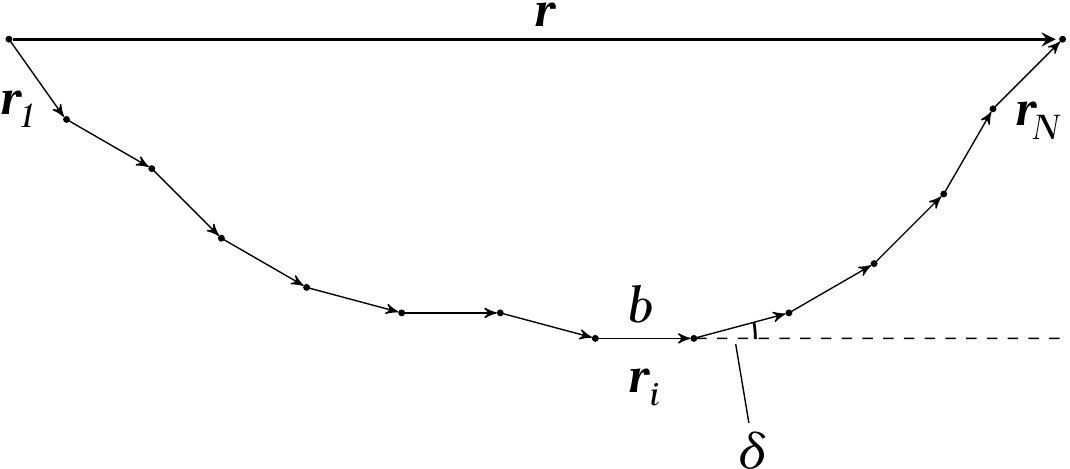}
}
\caption{Schematic representation of a Kratky--Porod chain.}
\label{KPChain}
\end{figure}

In turn, we can accordingly modify the definitions of the persistence length and other configurational parameters of DNA to include the effect of kinks and their influence on the state of the DNA coil. The Kratky--Porod model defines the persistence length of a macromolecular chain as a limit of the average value of the scalar product of the first segment unit vector $\vec{r}_1$ and the vector $\vec{r}$, which is the sum of the segment vectors $\vec{r}_i$ (see Fig.~\ref{KPChain})~\cite{cantor,hagerman},
\begin{eqnarray}
A_0 = \lim\limits_{N \rightarrow \infty} \langle \frac{\vec{r}_1}{b} \cdot \sum\limits_{i=1}^N \vec{r}_i \rangle = b \lim\limits_{N \rightarrow \infty} \sum\limits_{i=0}^{N-1} \langle \cos \delta \rangle ^i, \label{PS}
\end{eqnarray}
where $b$ is the length of each segment and $N$ is the number of segments. Taking the limit $N\rightarrow\infty$ and using the formula for a geometric progression we can obtain the following expression for the persistence length,
\begin{eqnarray}
A_0 = \frac{b}{1-\langle \cos\delta \rangle}.
\end{eqnarray}
Such a definition of the persistence length represents a chain with segments that are all in the normal state.

Let us start with \textit{type 1} kinks, which involve only a single segment. According to the definition of the persistence length~(\ref{PS}) a contribution of the $i$-th segment in a normal state is represented by the $\langle\cos\delta\rangle^i$ term in the series. On the other hand, if there is a probability $W_1$ of the type 1 kink formation on the first segment, then it should produce a contribution $W_1 \cos\theta\langle\cos\delta\rangle$ in a kinked state and a contribution $(1-W_1)\langle\cos\delta\rangle$ in a normal state to the series~(\ref{PS}). Accordingly, the total contribution to the persistence length produced by the first segment is $\bigl(1 - W_1 \bigl(1 - \cos\theta \bigr)\bigr) \langle \cos \delta\rangle$ in this case. If it is assumed that each segment of the chain can undergo the formation of a type 1 kink with a probability $W_1$, then the contributions of each segment to the persistence length should be modified in the following way~\cite{simonov},
\begin{eqnarray}
& \langle \cos \delta \rangle \rightarrow \bigl(1 - W_1 \bigl(1 - \cos\theta \bigr)\bigr) \langle \cos \delta \rangle, \\
& \nonumber \langle \cos \delta \rangle^2 \rightarrow \bigl(1 - W_1 \bigl(1 - \cos\theta \bigr)\bigr)^2 \langle \cos \delta \rangle^2, \\
& \nonumber ... \\
& \nonumber  \langle \cos \delta \rangle^i \rightarrow \bigl(1 - W_1 \bigl(1 - \cos\theta \bigr)\bigr)^i \langle \cos \delta \rangle^i, \\
& \nonumber ...
\end{eqnarray}
and, collecting the modified contributions of all segments, the persistence length of the chain with type 1 kinks can be calculated as
\begin{eqnarray}
\mathcal{A}_1 = \frac{b}{1- \bigl(1 - W_1 \bigl( 1- \cos\theta \bigr) \bigr) \langle \cos \delta \rangle}.
\end{eqnarray}

In this way, the presence of kinks in the DNA macromolecule decreases the persistence length (and hence the stiffness of the macromolecule) due to the new term $W_1(1-\cos\theta)\langle \cos\delta \rangle$ depending on the angle of the kink and the probability of its formation. Furthermore, the maximal decrease of the persistence length would be observed in the case of $\theta = 180^{\circ}$ due to the term $2W_1\langle \cos\delta \rangle$.

A type 2 kink is distributed over two base pair steps. Both segments are kinked by an equal angle, $\theta/2$, so the total kink angle is $\theta$. Therefore, if there is a probability $W_2$ of a type 2 kink formation on the first two segments, they should produce a contribution \
\begin{eqnarray}
\nonumber W_2 \Bigl(\cos\frac{\theta}{2}\langle\cos\delta\rangle + \cos^2\frac{\theta}{2} \langle\cos\delta\rangle^2\Bigr)
\end{eqnarray}
in a kinked state and a contribution 
\begin{eqnarray}
\nonumber (1-W_2)\Bigl(\langle\cos\delta\rangle + \langle\cos\delta\rangle^2\Bigr)
\end{eqnarray}
in a normal state to the series~(\ref{PS}). Hence, the total contribution to the persistence length produced by the first two segments is
\begin{eqnarray}
\nonumber \Bigl(1-W_2\Bigl(1-\cos \frac{\theta}{2} \Bigr)\Bigr) \langle \cos \delta \rangle + \\
\nonumber \Bigl(1-W_2\Bigl(1-\cos ^2 \frac{\theta}{2} \Bigr)\Bigr) \langle \cos \delta \rangle ^2
\end{eqnarray}
in this case. If it is assumed that each segment of the chain can undergo the formation of a type 2 kink with a probability $W_2$, then the contributions of each segment to the persistence length should be modified in the following way~\cite{simonov},
\begin{eqnarray}
\displaystyle
& \langle \cos \delta \rangle + \langle \cos \delta \rangle ^2 \rightarrow
\\
\nonumber & \bigl(1-W_2\bigl(1-\cos \frac{\theta}{2} \bigr)\bigr) \langle \cos \delta \rangle +
\\
\nonumber & \bigl(1-W_2\bigl(1-\cos ^2 \frac{\theta}{2} \bigr)\bigr) \langle \cos \delta \rangle ^2, \\\nonumber \\
& \nonumber \langle \cos \delta \rangle^3 + \langle \cos \delta \rangle ^4 \rightarrow
\\
\nonumber & \bigl(1-W_2\bigl(1-\cos ^2 \frac{\theta}{2} \bigr)\bigr)\bigl(1-W_2\bigl(1-\cos \frac{\theta}{2} \bigr)\bigr) \langle \cos \delta \rangle^3 +
\\
\nonumber & \bigl(1-W_2\bigl(1-\cos ^2 \frac{\theta}{2} \bigr)\bigr)^2 \langle \cos \delta \rangle ^4, \\
& \nonumber ... \\
& \nonumber \langle \cos \delta \rangle^{2i} + \langle \cos \delta \rangle ^{2i+1} \rightarrow \\
\nonumber & \bigl(1-W_2\bigl(1-\cos ^2 \frac{\theta}{2} \bigr)\bigr)^i \bigl(1-W_2\bigl(1-\cos \frac{\theta}{2} \bigr)\bigr) \langle \cos \delta \rangle^{2i} +
\\
\nonumber & \bigl(1-W_2\bigl(1-\cos ^2 \frac{\theta}{2} \bigr)\bigr)^{i+1} \langle \cos \delta \rangle ^{2i+1}, \\
& \nonumber ...
\end{eqnarray}
Summing up all the contributions, we obtain the following expression for the persistence length of a chain with type 2 kinks,
\begin{eqnarray}
\mathcal{A}_{2} = \frac{b \cdot F_2 \bigl(W_2, \theta \bigr)}{1- \bigl(1 - W_2 \bigl( 1- \cos ^2 \frac{\theta}{2} \bigr) \bigr) \langle \cos \delta \rangle ^2},
\end{eqnarray}
where $F_2 \bigl(W_2, \theta \bigr) = 1 + \bigl(1-W_2\bigl(1-\cos \frac{\theta}{2} \bigr)\bigr) \langle \cos \delta \rangle$ characterises the two segments involved by a kink of type 2. In this way, it can be seen that kinks of type 2 should decrease the stiffness of the macromolecule more smoothly in contrast to kinks of type 1.

In the previous section it was assumed that the so-called conformational kinks involving a more complicated structure than kinks of types 1 and 2, could also exist. In this way, the obtained approach can be generalised by focusing on the kinks with an arbitrary distribution of angles $\theta_1, \theta_2, ..., \theta_n$ inside the kink and involving $n\geq 3$ base pair steps. The probability of the formation of a conformational kink ($W_n$) should be introduced, and the contribution of $n$ chain segments undergoing a conformational kink can be changed in the same way as for the kinks of types 1 and 2~\cite{simonov},
\begin{eqnarray}
\displaystyle
& \langle \cos \delta \rangle + \langle \cos \delta \rangle ^2 + ... + \langle \cos \delta \rangle ^n \rightarrow \\
\nonumber & \bigl(1-W_n\bigl(1-\cos \theta_1 \bigr)\bigr) \langle \cos \delta \rangle + \\
\nonumber & \bigl(1-W_n\bigl(1-\cos \theta_1 \cdot \cos \theta_2 \bigr)\bigr) \langle \cos \delta \rangle ^2 + ... + \\
\nonumber & \bigl(1-W_n\bigl(1-\cos \theta_1 \cdot \cos \theta_2 \cdot ... \cdot \cos \theta_n \bigr)\bigr) \langle \cos \delta \rangle ^n = \\
\nonumber & \sum\limits_{j=1}^{n} \bigl(1 - W_n \bigl( 1- \prod\limits_{k=1}^j \cos \theta_k \bigr) \bigr) \langle \cos \delta \rangle ^j , \\\nonumber \\
& \nonumber \langle \cos \delta \rangle^{n+1} + \langle \cos \delta \rangle ^{n+2} + ... + \langle \cos \delta \rangle^{2n} \rightarrow \\
\nonumber & \bigl(1 - W_n \bigl( 1- \prod\limits_{k=1}^n \cos \theta_k \bigr) \bigr)  \cdot \\
& \nonumber \sum\limits_{j=1}^{n} \bigl(1 - W_n \bigl( 1- \prod\limits_{k=1}^j \cos \theta_k \bigr) \bigr) \langle \cos \delta \rangle ^j, \\
& \nonumber ... \\
& \nonumber \langle \cos \delta \rangle^{in+1} + \langle \cos \delta \rangle ^{in+2} + ... + \langle \cos \delta \rangle^{(i+1)n} \rightarrow \\
\nonumber & \bigl(1 - W_n \bigl( 1- \prod\limits_{k=1}^n \cos \theta_k \bigr) \bigr)^i  \cdot \\
& \nonumber \sum\limits_{j=1}^{n} \bigl(1 - W_n \bigl( 1- \prod\limits_{k=1}^j \cos \theta_k \bigr) \bigr) \langle \cos \delta \rangle ^j, \\
& \nonumber ...
\end{eqnarray}
Accordingly, the persistence length of a chain with such kinks can be defined as
\begin{eqnarray}
\mathcal{A}_{n} = \frac{b \cdot F_n \bigl(W_n, \{\theta_k\} \bigr)}{1- \bigl(1 - W_n \bigl( 1- \prod\limits_{k=1}^{n} \cos \theta_k \bigr) \bigr) \langle \cos \delta \rangle ^{n}},
\end{eqnarray}
where

$F_n \bigl(W_n, \{\theta_k\} \bigr) = 1 + \sum\limits_{i=1}^{n-1} \bigl(1 - W_n \bigl( 1- \prod\limits_{k=1}^i \cos \theta_k \bigr) \bigr) \langle \cos \delta \rangle ^i$ characterises segments of the chain modified by the conformational kink, and $\theta_k$ is the angle between the $k$-th and $(k+1)$-th base pair.

Coil size and gyration radius of the DNA chain are the simplest parameters which characterise its spatial dimensions and can describe the properties of the double helix in solution. In the case of a kinked DNA these parameters can be calculated exactly. It is only necessary to replace the persistence length of the undisturbed DNA with that of the kinked DNA in the corresponding expressions. Thus, the coil size of a chain with kinks of the chosen type can be defined as
\begin{eqnarray}
\langle \mathcal{R}_{n} \rangle^2 = [\langle R \rangle^2]_{A\rightarrow \mathcal{A}_n} = 2 \mathcal{A}_n^2 \Bigl(\frac{L}{\mathcal{A}_n} - 1 + e^{-L/\mathcal{A}_n}\Bigr),
\end{eqnarray}
where $\mathcal{A}_n$ is the persistence length of the chain with kinks of a chosen type and $L$ is its contour length. In the same way, the gyration radius can be defined as
\begin{eqnarray}
& \langle \mathcal{G}_{n} \rangle^2 = [\langle G \rangle^2]_{A\rightarrow \mathcal{A}_n} = 
\\
\nonumber & \frac{L\mathcal{A}_n}{3} - \mathcal{A}_n^2 + \frac{2\mathcal{A}_n^3}{L} - \frac{2\mathcal{A}_n^4}{L^2} (1 - e^{-L/\mathcal{A}_n}).
\end{eqnarray}
Consequently, the coil size and the gyration radius experience the same renormalisation due to the presence of kinks and hold the same expressions as the ones of the KWLC model~\cite{wiggins}. On the other hand, the persistence length is calculated differently than the KWLC model does, since the formation of kinks of different types is taken into account.

\section{Bending of DNA by multivalent cations}
\label{sec:Cations}

In the preceding section we obtained an approach for the Kratky--Porod model that describes the persistence length of chains with kinks characterised by two parameters, kink angle ($\theta$) and kink formation probability ($W$). Let us begin a discussion of the results by calculating the persistence length of chains with three types of kinks: type 1, type 2 and conformational kinks distributed over 3 helix steps. We assume that all the segments inside a conformational kink have equal bending angles $\theta_k = \theta / 3$. Furthermore, we take $A_0=500$\r{A} as the persistence length of an unperturbed double helix and $b=3.4$\r{A} as the length of a chain segment.

\begin{figure}\resizebox{0.5\textwidth}{!}{%
  \includegraphics{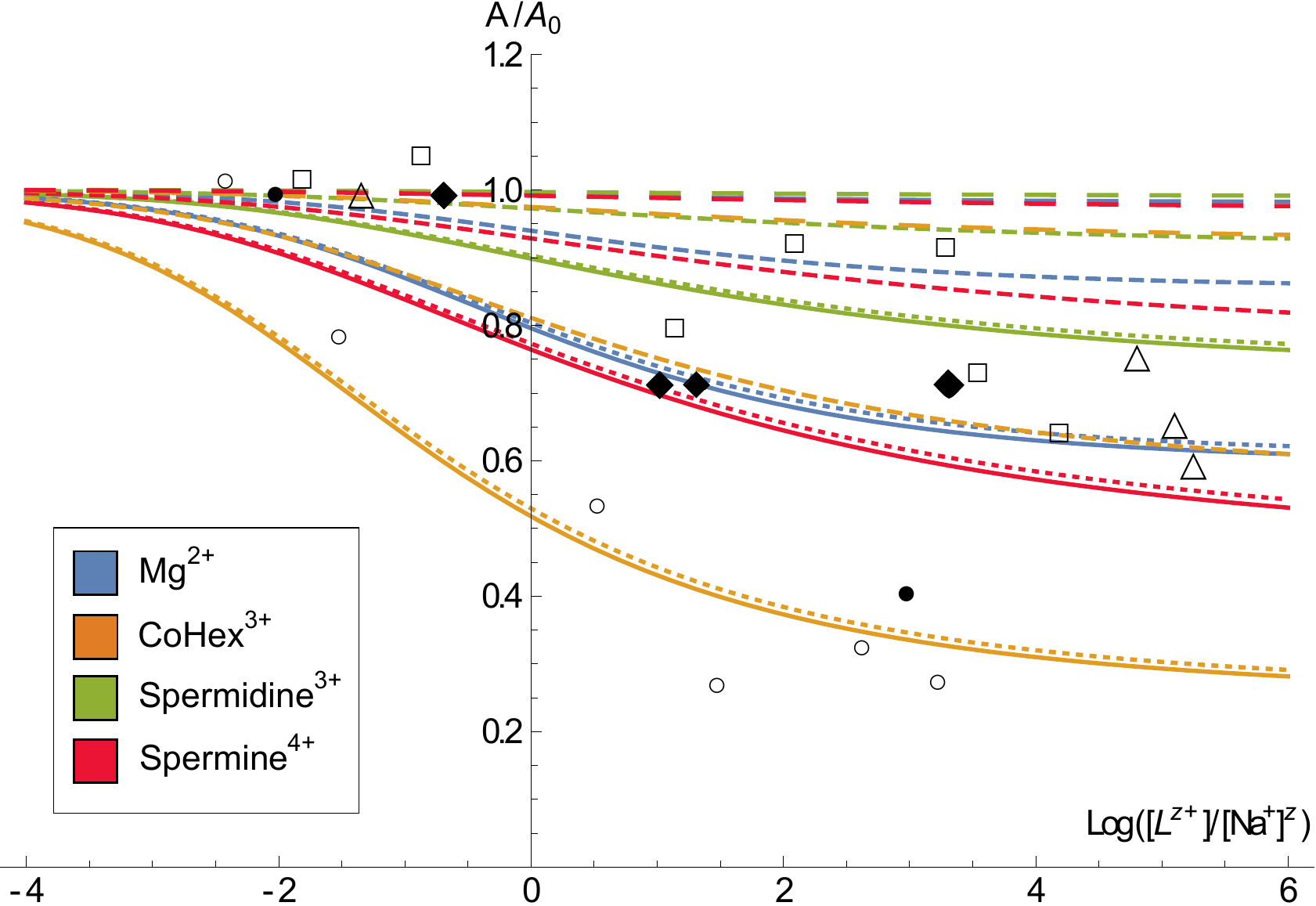}
}\caption{(Color online) The reduction of the persistence length $A$ is caused by the presence of multivalent cations. The experimental points are reproduced from the paper by Rouzina and Bloomfield~\cite{rouzina} and represent the presence of the following multivalent cations: diamonds, \textit{Mg}$^{2+}$~\cite{porschke}; empty circles, \textit{CoHex}$^{3+}$~\cite{baumann,baumannunpubl}; filled circles, \textit{CoHex}$^{3+}$~\cite{porschke}; squares, \textit{Spermidine}$^{3+}$~\cite{baumann,baumannunpubl}; triangles, \textit{Spermine}$^{4+}$~\cite{porschke}. The curves represent relative persistence lengths of B-DNA in solution with corresponding cations. Solid curves correspond to kinks of type 1, short-dashed curves correspond to kinks of type 2, long-dashed curves correspond to conformational kinks distributed over 6 base pairs and dotted curves correspond to the Rouzina--Bloomfield persistence length $\mathcal{A}_{RB}/A_0$.}\label{Results_Concentration}\end{figure}
We focus on the bending of DNA by multivalent cations as a good candidate for applying the obtained model. As shown by Rouzina and Bloomfield, the persistence length can be dramatically reduced due to the presence of small multivalent cations in solution~\cite{rouzina}. They proposed to extend the Kratky--Porod model to take into account the bending induced by cations: each cation causes a bend of a small angle $\beta_i$, that is actually statistically independent of the bending in the absence of cations. Thus, they define the persistence length in the following form,
\begin{eqnarray}
\mathcal{A}_{RB} = \frac{A_0}{1 + W_i \langle \beta_i^2 \rangle / \langle \beta_0^2 \rangle},
\end{eqnarray}
where $A_0=500$\r{A}, $W_i$ is the probability of bending and $(\langle\beta_0^2\rangle)^{1/2} = 6.7^{\circ}$. On the other hand, Rouzina and Bloomfield suggested to also consider the bends induced by ca\-tions as distributed over 6 steps due to the proposed electrostatic bending mechanism~\cite{rouzina,baumann}. Generally speaking, we can imagine such bends as conformational kinks with $n=6$ and $\theta_k = \theta / 6$, where $\theta$ is the angle of the whole kink.

The probability of the formation of a bend is suggested to be proportional to the number of cations bound per base pair $\Theta_z$, so $W_i = W_{i0} \cdot \Theta_z$, where the fractional occupancy $\Theta_z$ can be estimated by solving the equation~\cite{rouzina,rouzina1997}
\begin{eqnarray}
\frac{z\Theta_z}{2} = \frac{[\mbox{\textit{L}}^{z+}]}{[\mbox{\textit{Na}}^{+}]^z} n_s^{z-1} \Bigl( 1 - \frac{z\Theta_z}{2} \Bigr)^z,
\end{eqnarray}
where $z$ is a charge, $n_s$ is the concentration of the cation on the DNA surface, $[\mbox{\textit{L}}^{z+}]$ and $[\mbox{\textit{Na}}^{+}]$ are bulk concentrations of the ligand and $\mbox{\textit{Na}}^{+}$ respectively. Rouzina and Bloomfield assumed that every bound cation produces an equivalent bend, and hence $W_{i0} = 2/z$.

\begin{figure}\resizebox{0.5\textwidth}{!}{%
  \includegraphics{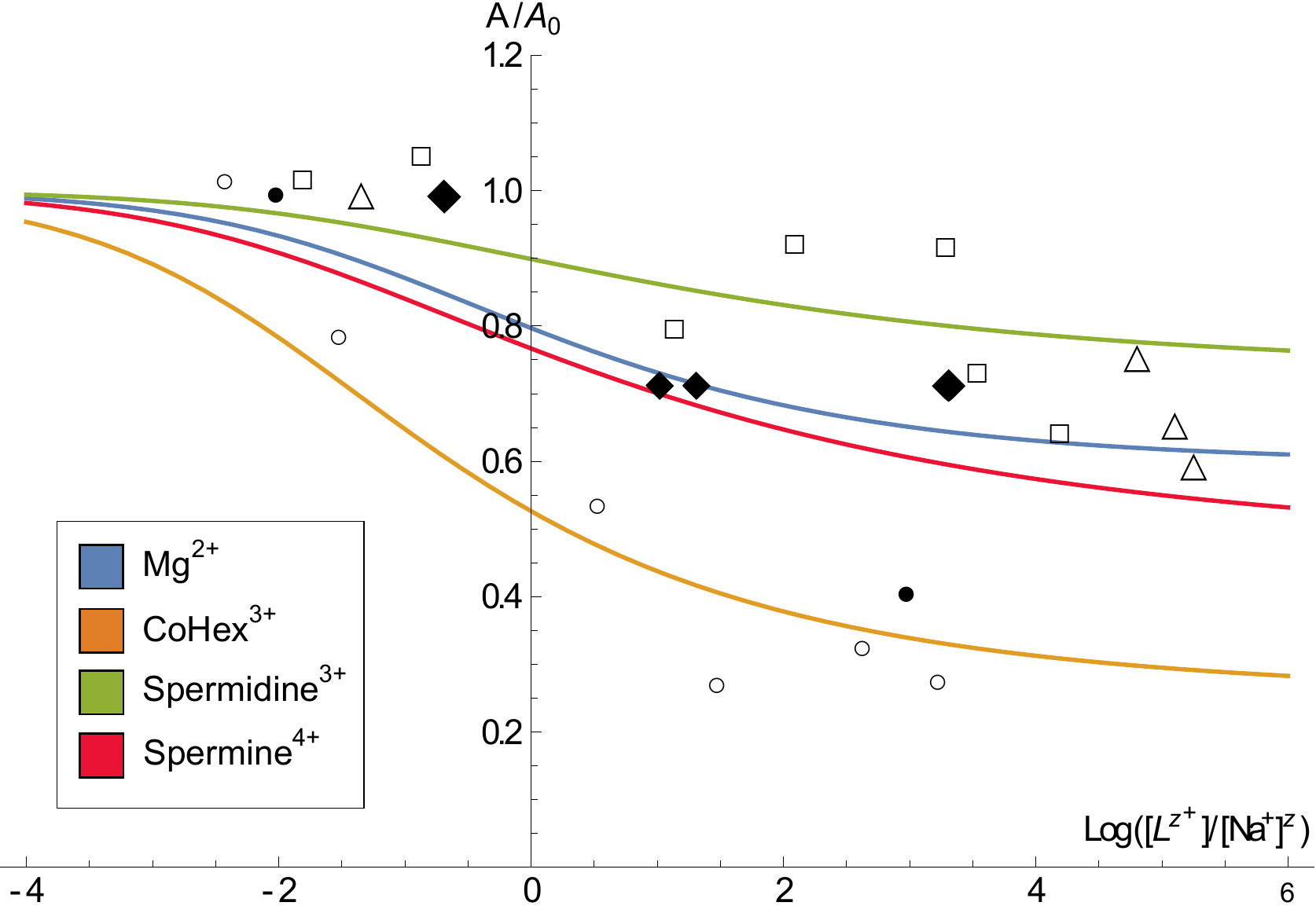}
}\caption{(Color online) Reduction of the persistence length $A$ due to conformational kinks distributed over 6 steps with the changed parameters caused by the presence of multivalent cations. The experimental points are reproduced from the paper by Rouzina and Bloomfield~\cite{rouzina} and represent the presence of the following multivalent cations: diamonds, \textit{Mg}$^{2+}$~\cite{porschke}; empty circles, \textit{CoHex}$^{3+}$~\cite{baumann,baumannunpubl}; filled circles, \textit{CoHex}$^{3+}$~\cite{porschke}; squares, \textit{Spermidine}$^{3+}$~\cite{baumann,baumannunpubl}; triangles, \textit{Spermine}$^{4+}$~\cite{porschke}. The curves represent relative persistences length of B-DNA in solution with corresponding cations.}\label{Results_Concentration_Conformational}\end{figure}

In this way, we can compare results for kinks and Rouzina--Bloomfield bends with the experimental data obtained for \textit{Mg}$^{2+}$, \textit{CoHex}$^{3+}$, \textit{Spermidine}$^{3+}$ and \textit{Spermine}$^{4+}$ (see Fig.~\ref{Results_Concentration}). We assume that each bound cation causes an equivalent bend and we use the same formula to obtain the number of cations per base pair. Furthermore, we choose $n_s = 1 M$ and the same angles for kinks of type 1 and type 2 as Rouzina and Bloomfield do for bends, $5.5^{\circ}$ for \textit{Mg}$^{2+}$, $11.5^{\circ}$ for \textit{CoHex}$^{3+}$, $4.0^{\circ}$ for \textit{Spermidine}$^{3+}$ and $7.1^{\circ}$ for \textit{Spermine}$^{4+}$~\cite{rouzina}. For conformational kinks distributed over 6 steps we assume the same parameters.

In Fig.~\ref{Results_Concentration}, we compare the results for the persistence length of DNA with kinks of type 1 (solid curves), type 2 (short-dashed curves) and conformational kinks distribu\-ted over 6 steps (long-dashed curves) against the Rouzina--Bloom\-field persistence length (dotted curves) and experimental points. According to the obtained results the reduction of the persistence length by Rouzina--Bloomfield bends is highly compatible with a corresponding reduction caused by kinks of type 1. However, this fact is not surprising since $\mathcal{A}_{RB}$ can be obtained from $\mathcal{A}_1$ using a Taylor expansion in powers of the kink angle. As we expect, kinks of type 1 cause a dramatic decrease of the persistence length. Particularly, our approach predicts that the persistence length decreases by $60$-$70\%$ for high concentrations of \textit{CoHex}$^{3+}$, $35$-$40\%$ for \textit{Spermine}$^{4+}$, $30$-$35\%$ for \textit{Mg}$^{2+}$ and $20\%$ for \textit{Spermidine}$^{3+}$. Moreover, predictions for kinks of type 1 are compatible with the experimental data in the range of high concentrations of the ligand, especially for the \textit{CoHex}$^{3+}$ solutions. On the other hand, if we assume the presence of kinks of type 2 caused by multivalent cations, we will observe a much smaller decrease of the persistence length, such as $25$-$35\%$ for \textit{CoHex}$^{3+}$ in particular. Conformational kinks distributed over 6 steps do not decrease the persistence length sufficiently for all considered multivalent cations.

Consequently, if we assume the same parameters of kinks as Rouzina and Bloomfield use for bends and compare the obtained results with the experimental data, we can see that the decrease of the persistence length due to the presence of multivalent cations can be adequately described by type 1 kinks, at least for relatively high concentrations of the ligand. Type 2 kinks and especially conformational kinks distributed over 6 steps with such parameters do not cause the same decrease of the persistence length as observed experimentally. Therefore, we can conclude that the bending of DNA can be adequately described with type 1 kinks only and not by conformational kinks distributed over 6 steps if we assume that the values of the kink angle and the formation probability are equal to those in~\cite{rouzina}.

However, we can change the parameters of a kink to make the approach consistent in the case of kinks of other types. For example, presented in Fig.~\ref{Results_Concentration_Conformational} are the results of the persistence length of DNA with conformational kinks distributed over 6 steps. However, we assume that the angle of each kink is increased now and choose $33^{\circ}$ for \textit{Mg}$^{2+}$, $69^{\circ}$ for \textit{CoHex}$^{3+}$, $24^{\circ}$ for \textit{Spermidine}$^{3+}$ and $42.6^{\circ}$ for \textit{Spermine}$^{4+}$. We can see that the kinks modified in such a way can also adequately describe the decrease of the persistence length in the presence of multivalent cations. In this way, we have to conclude that experimental results can be explained using not only kinks of type 1 (that are used in the Rouzina--Bloomfield model in fact) but kinks of other types as well, particularly conformational kinks over 6 steps with the angles fixed in Fig.~\ref{Results_Concentration_Conformational}. Therefore, more experimental data is needed. In particular, it is necessary to measure the angles of kinks that occur in the double helix due to the action of multivalent cations to identify their type and configuration adequately. Hence, when the angle and formation probability of a kink are known, it is possible to predict its type and the value of decrease of the persistence length caused by the presence of such kinks.

\section{Summary and outlook}
\label{sec:Summary}
In this paper we have reviewed a simple approach for the Kratky--Porod model presented in~\cite{simonov}, which allows the calculation of the persistence length of the DNA double helix with kinks of a certain type occurring in its structure. This approach focuses on two basic types of kinks proposed and observed in the computer simulations by Lanka\v{s} and colleagues~\cite{prevost,lankas}, and kinks with more complex structure called conformational kinks. In that way, corresponding configurational parameters of DNA including the persistence length can be measured and compared with the predictions of the model. Our approach uses two parameters, the total kink angle ($\theta$) and the probability of kink formation ($W$). Consequently, it describes the flexibility of a DNA chain with kinks of an arbitrary intrinsic structure and length $n$, contrasting with the KWLC model, which describes the kinks of type 1 only~\cite{wiggins}.

The analysis of the approach showed that the possibility of the formation of kinks in the double helix dramatically reduces its persistence length. In particular, it undergoes the strongest decrease due to type 1 kinks. On the other hand, changes in the nature of kinks (herewith, its geometry and type) can decrease the persistence length as well as increase, in comparison with the persistence length of the double helix with type 1 kinks. Thus, it is possible to determine the concentration of kinks in the DNA chain and their nature by an analysis of the predicted configurational parameters.

Bending of DNA by multivalent cations provides a good example to apply the discussed approach for practical computations of the persistence length. Particularly, the model developed by Rouzina and Bloomfield~\cite{rouzina} perfectly agrees with our approach in the case of type 1 kinks, such as single-stranded breaks. However, the decrease of the persistence length in the case of the presence of multivalent cations can be described not only by single-stranded breaks, but also by the kinks with another intrinsic structure. For example, results for the persistence length of the double helix with conformational kinks distributed over 6 base pairs are also compatible with results of the Rouzina-Bloomfield framework. But the parameters of a kink, in particular the bending angle, are much higher and provide a stronger bending of the double helix in this case. Therefore, an experiment to determine the nature of the occurred kinks, their angles and intrinsic structure is strongly needed.

Furthermore, there is an important challenge to extend the proposed approach to more complicated configurations of the DNA macromolecule, e.~g. with kinks of different types in its structure. In particular, configurations of the DNA double helix containing both type 1 and type 2 kinks were observed in the simulations of DNA minicircles~\cite{lankas,mitchell}. Another challenge is to take into account the sequence dependence of the persistence length, which is a significant factor in DNA-protein interaction~\cite{geggier}. Such an improved framework would be a more realistic theoretical tool for the analysis of the flexibility and conformational properties of the kinkable DNA double helix involved in key biological processes such as folding, transcription and others.

\textbf{Acknowledgment.} The author thanks Sergey N. Vol\-kov for a significant contribution to this work and Sergiy Perepelytsya and Polina Kanevska (Bogolyubov Institute for Theoretical Physics of the NAS of Ukraine) for the fruitful discussions.

\end{document}